\documentclass[prd,showpacs]{revtex4}
\begin{document}
\title{
Finite temperature Casimir effect for massless Majorana fermions in a magnetic field
      }
\author{Andrea Erdas}
\affiliation{
 Department of Physics, Loyola University Maryland, 4501 North Charles Street
       Baltimore, Maryland 21210, USA}
\email{aerdas@loyola.edu}
\begin {abstract} 
The zeta function regularization technique is used to study the finite
temperature Casimir effect
for a massless Majorana fermion field confined between parallel plates and satisfying bag boundary
conditions. A magnetic field perpendicular to the plates is included. An expression 
for the zeta function is obtained, which is exact to all orders in the magnetic 
field strength, temperature and plate distance. The zeta function is used to calculate 
the Helmholtz free energy of the Majorana field and the pressure on the plates, in the case of weak magnetic field and strong
magnetic field. In both cases, simple analytic expressions are obtained for the free energy and pressure
which are very accurate and valid for all values of the temperature and plate distance.
\end {abstract}
\maketitle
\section{Introduction}
\label{1}
The Casimir effect is a purely quantum phenomena where an attractive or repulsive force occurs between 
electrically neutral conducting plates in vacuum. The effect was first discovered by Casimir \cite{Casimir:1948dh},
who calculated the attractive electromagnetic force between two parallel conducting plates, and can be regarded as a quantitative proof
of the quantum fluctuations of the electromagnetic field. The repulsive Casimir effect was discovered by Boyer some time later \cite{Boyer:1968uf}, 
when he showed that if the electromagnetic field is confined 
inside a perfectly conducting sphere, the wall of the sphere is subject to a repulsive force. 
The first experimental evidence of the Casimir force was obtained more than fifty years ago by Sparnaay \cite{Sparnaay:1958wg}
and, since then, many greatly improved experimental observations of this effect have been reported. For a comprehensive review 
of these experiments see the review article by Bordag et al. \cite{Bordag:2001qi}.

Since Casimir forces have many applications, from nanotubes and nanotechnology \cite{Bellucci:2009jr,Bellucci:2009hh}, 
to branes and compactified extra dimensions \cite{Poppenhaeger,Cheng,Cheng2,
Hertzberg,Hertzberg2,Marachevsky:2007da,Edery,Edery2,Kirsten,Milton,Elizalde,Elizalde2,Flachi,Flachi2,Garriga,
Frank,Linares}
and to string theory \cite{Saharian,Hadasz},
a great deal of effort has gone into studying the Casimir effect
and its generalization to quantum fields other than the electromagnetic: fermions, bosons and other scalar fields
have all been investigated extensively \cite{Bordag:2001qi}. 

It is well known that the Casimir effect
is very sensitive to the boundary conditions for all types of quantum fields, and the most used ones
are Dirichlet and Neuman boundary conditions on the plates. These boundary conditions however cannot be used in the case
of fermion fields or fields with spin in general \cite{Ambjorn:1981xw} and therefore, for fermions,
the bag boundary conditions are used, which originally were introduced to find a solution to confinement \cite{Chodos:1974je}.
In this work I will use the bag boundary conditions for a Majorana fermion field confined between two parallel
plates. 

Majorana fermions appear in many different areas of physics: in some models of extra dimensions a Majorana fermion,
the Kaluza-Klein neutrino, is a leading WIMP candidate \cite{Oikonomou:2006mh}. In the field of superconductivity, a Majorana
bound state is theoretically predicted in rotating superfluid $^3$He - A between parallel plates and in 
the presence of a magnetic field \cite{tsutsumi}. While the Casimir effect for Majorana fermion
fields has been studied in vacuum \cite{DePaola:1999im,Oikonomou:2009zr} and recently at finite temperature \cite{Cheng:2010kc}, a study that
considers magnetic effects in vacuum and thermal and magnetic effects combined has not appeared in the literature. 
This paper will therefore investigate the Casimir effect for Majorana fermion fields at finite temperature 
and in the presence of a magnetic field.

Casimir effect calculations must address carefully the issue of the vacuum energy regularization. Many regularization techniques 
are available nowadays, and two of them have been applied frequently and successfully to the Casimir effect: the cutoff method 
often used in various piston configurations
\cite{Cavalcanti:2003tw,Oikonomou:2009zr} and the zeta function technique \cite{Elizalde:2007du,Elizalde:2006iu,Elizalde:1988rh}. 
My choice for this paper is the zeta function technique, which is very
powerful and is also used in the computation of effective actions \cite{dittrich,erdasfeld}. I will apply this regularization method
to calculate the free energy and pressure on the plates for a Majorana fermion field with bag boundary conditions
confined between two parallel plates, at a distance $a$ from each other. I assume this system to be in thermal equilibrium 
with a heat reservoir at finite temperature $T$, and will use the imaginary time 
formalism of finite temperature field theory, which is suitable for a system in thermal equilibrium. 
A magnetic field $\vec B$ is present in the region between the plates and is perpendicular
to the plates.

In Section \ref{2}, I calculate the zeta function for this system, exact to all orders in $eB$, $T$ and $a$, where 
$e$ is the Majorana field charge. 
In Section \ref{3}, I use the zeta function obtained in the previous section, to calculate the Helmholtz free energy 
of the Majorana fermion field and the pressure on the plates and obtain simple analytic expressions for these quantities
in the case of weak magnetic field ($eB\ll T^2, a^{-2}$), and of strong magnetic field ($eB\gg T^2, a^{-2}$).
A discussion of my results is presented in Section \ref{4}. 

\section{ Zeta function evaluation}
\label{2}

Using the imaginary time 
formalism of finite temperature field theory I write
the partition function ${\cal Z}$ for a fermionic system in thermal equilibrium at finite temperature $T$
\begin{equation}
{\cal Z}=N\int_{\rm Antiperiodic}\!\!\!\!\!\!\!\!\!\!\!\!\!\!\!\!\!\!\!\!\!\!
D{\bar \psi} \,\,D\psi\,\, \exp\left(\int_0^\beta d\tau\int d^3x{\cal L}\right)
\label{partition}
\end{equation}
where ${\cal L}$ is the Lagrangian density for the fermionic system, $N$ is a constant and 'antiperiodic' means 
that this functional integral is evaluated over field configurations satisfying
\begin{equation}
\psi(x,y,z,0)=-\psi(x,y,z,\beta),
\label{antiperiodic}
\end{equation}
where $\beta=1/T$ is the antiperiodic length in the Euclidean time axis. In addition to the boundary conditions
given by (\ref{antiperiodic}), I impose bag boundary conditions \cite{Chodos:1974je} for Majorana fermions between two plates.
In 3-dimensional space with two very large parallel plates perpendicular to the $z$-axis
and located at $z=0$ and $z=a$, the MIT bag boundary conditions constrain the Majorana fermion fields
to the region between the plates. These boundary conditions
are expressed in covariant form as 
\begin{equation}
n^\mu{\bar\psi}\gamma_\mu\psi=0
\label{bag}
\end{equation}
where $n^\mu = (0, {\vec n})$ and ${\vec n}$ is the vector normal to the surface of the plates and directed towards the 
interior of the slab region. The above boundary conditions prevent the flow of fermion current out of the slab region.
In the slab region there is also a uniform magnetic field pointing in the $z$ direction, ${\vec B}=(0,0,B)$. The Majorana fields
have charge $e$ and will interact with the magnetic field.

The Helmholtz free energy $F$ and the partition function ${\cal Z}$ are related by 
\begin{equation}
F=-\beta^{-1}\log {\cal Z}, 
\label{F}
\end{equation}
therefore I
evaluate the functional integral (\ref{partition}) and, after the straightforward gamma algebra, find
\begin{equation}
2 \log {\cal Z}=\log \,\det \left(-D^{(-)}_{\rm E}|{\cal F}_a\right) 
+ \log \,\det \left(-D^{(+)}_{\rm E}|{\cal F}_a\right)
\label{logZ}
\end{equation}
where the symbol ${\cal F}_a$ indicates the set of functions which satisfy boundary conditions 
(\ref{antiperiodic}) and (\ref{bag}), and the operators $D^{(\pm )}_{\rm E}$ are defined as:
\begin{equation}
D^{(\pm )}_{\rm E} = \partial^2_\tau+\partial^2_z-({\vec p} -e{\vec A})^2_\perp\mp eB,
\label{D}
\end{equation}
where $\vec A$ is the electromagnetic vector potential, the subscript E indicates Euclidean time, and I used 
the notation ${\vec p}_\perp=(p_x,p_y,0)$.

The zeta function technique allows me to use the eigenvalues of $-D^{(\pm )}_{\rm E}$ to evaluate $\log {\cal Z}$.
The bag boundary conditions (\ref{bag}) are satisfied only
if the allowed values for the momentum in the $z$-direction are
\begin{equation}
p_z={\pi\over a}\left(l+{1\over 2}\right),
\label{pz}
\end{equation}
where $l \in \{0, 1, 2, 3,...\}$, and therefore
the eigenvalues of $-\partial^2_\tau-\partial^2_z$ whose eigenfunctions satisfy (\ref{antiperiodic}) and (\ref{bag}) are:
\begin{equation}
{\pi^2\over a^2}\left(l+{1\over 2}\right)^2+{4\pi^2\over\beta^2}\left(m+{1\over 2}\right)^2,
\label{eigenvalues1}
\end{equation}
where $l \in \{0, 1, 2, 3,...\}$ and $m \in \{0, \pm 1, \pm 2, \pm 3,...\}$. The spectrum of the operator 
$({\vec p} -e{\vec A})^2_\perp$ is well known from one-particle quantum mechanics, and its eigenvalues are
\begin{equation}
2eB\left(n+{1\over 2}\right),
\label{eigenvalues2}
\end{equation}
with $n \in \{0, 1, 2, 3,...\}$. Using the eigenvalues (\ref{eigenvalues1}) and (\ref{eigenvalues2}), I construct
the zeta functions $\zeta\left(s,-D^{(+)}_{\rm E}\right)$ and $\zeta\left(s,-D^{(-)}_{\rm E}\right)$, which are given by
\begin{equation}
\zeta\left(s,-D^{(\pm )}_{\rm E}\right)=L^2 \sum_{l=0}^\infty \sum_{m=-\infty}^\infty \left(
{eB \over 2\pi}\right) \sum_{n=0}^\infty\left[
{\pi^2\over a^2}\left(l+{1\over 2}\right)^2+{4\pi^2\over\beta^2}\left(m+{1\over 2}\right)^2+2eB\left(n+{1\over 2}\right)
\pm eB\right]^{-s},
\label{zetapm}
\end{equation}
where $L^2$ is the area of the plates. I add them to obtain
\begin{equation}
\zeta(s)=\zeta\left(s,-D^{(+)}_{\rm E}\right)+\zeta\left(s,-D^{(-)}_{\rm E}\right),
\label{zeta}
\end{equation}
and find immediately, using the zeta function technique
\begin{equation}
2 \log {\cal Z}=-\zeta'(0).
\label{zandzeta}
\end{equation}
Using the following identity
\begin{equation}
z^{-s}={1\over \Gamma(s)}\int_0^\infty dt\, t^{s-1}e^{-zt},
\label{gamma}
\end{equation}
I rewrite $\zeta(s)$ as
\begin{equation}
\zeta(s)= {L^2 \over 2\pi}{1\over \Gamma(s)}\sum_{l=0}^\infty \sum_{m=-\infty}^\infty 
\int_0^\infty dt \, t^{s-2} eBt \coth eBt
\exp\left[-
{\pi^2\over a^2}\left(l+{1\over 2}\right)^2t-{4\pi^2\over\beta^2}\left(m+{1\over 2}\right)^2t\right],
\label{zeta2}
\end{equation}
where I also used
\begin{equation}
\sum_{n=0}^\infty\left[e^{-2nz}+e^{-2(n+1)z}\right]=\coth z.
\label{coth}
\end{equation}
After applying the Poisson resummation formula \cite{Dittrich:1979ux}
to the two sums, I obtain
\begin{equation}
\zeta(s)= {V \over 2\pi^2}{\beta\over \Gamma(s)}
\int_0^\infty dt \, t^{s-3} eBt \coth eBt
\left({1\over 2}+\sum_{l=1}^\infty (-1)^l e^{-a^2 l^2/t}\right)
\left({1\over 2}+\sum_{m=1}^\infty (-1)^m e^{-\beta^2 m^2/4t}\right),
\label{zeta3}
\end{equation}
where $V=L^2a$ is the volume of the slab.
It is evident from (\ref{zeta3}) that the Poisson resummation formula allows me to naturally divide $\zeta$ into four parts
\begin{equation}
\zeta(s)= \zeta_{HE,0}(s)+\zeta_{HE,\beta}(s)+\zeta_{C,0}(s)+\zeta_{C,\beta}(s),
\label{zeta4}
\end{equation}
two of which are vacuum parts that do not depend on $\beta$ except for the overall multiplicative factor, while the other two
are finite temperature corrections that vanish when $\beta\rightarrow\infty$. The four parts of $\zeta(s)$ are:
\begin{equation}
\zeta_{HE,0}(s)={L^2 \over 8\pi^2}{a\beta\over \Gamma(s)}
\int_0^\infty dt \, t^{s-3} eBt \coth eBt,
\label{zetae0}
\end{equation}
\begin{equation}
\zeta_{HE,\beta}(s)={L^2 \over 4\pi^2}{a\beta\over \Gamma(s)}
\int_0^\infty dt \, t^{s-3} eBt \coth eBt
\left(\sum_{m=1}^\infty (-1)^m e^{-\beta^2 m^2/4t}\right),
\label{zetaeb}
\end{equation}
\begin{equation}
\zeta_{C,0}(s)={L^2 \over 4\pi^2}{a\beta\over \Gamma(s)}
\int_0^\infty dt \, t^{s-3} eBt \coth eBt
\left(\sum_{l=1}^\infty (-1)^l e^{-a^2 l^2/t}\right),
\label{zetac0}
\end{equation}
\begin{equation}
\zeta_{C,\beta}(s)={L^2 \over 2\pi^2}{a\beta\over \Gamma(s)}
\int_0^\infty dt \, t^{s-3} eBt \coth eBt
\left(\sum_{l=1}^\infty (-1)^l e^{-a^2 l^2/t}\right)
\left(\sum_{m=1}^\infty (-1)^m e^{-\beta^2 m^2/4t}\right),
\label{zetacb}
\end{equation}
and each one will contribute differently to the free energy.
I will show in the next section that the contribution of $\zeta_{HE,0}$ to the free energy is the opposite of the 
unrenormalized Heisenberg-Euler effective lagrangian \cite{heisenberg} multiplied by the volume of the slab $V$, 
while the contribution 
of $\zeta_{HE,\beta}$ is the opposite of the finite temperature correction to the  Heisenberg-Euler 
effective lagrangian, also multiplied by $V$. The two remaining pieces, $\zeta_{C,0}$ 
and $\zeta_{C,\beta}$, will be shown to be the main contributors to the vacuum  Casimir energy and its finite temperature correction.
\section{ Free energy and pressure}
\label{3}
I combine (\ref{F}) and (\ref{zandzeta}) to obtain the free energy $F$ in terms of the zeta function
\begin{equation}
F={1\over 2\beta}\zeta'(0).
\label{Fandzeta}
\end{equation}
The derivatives of the four parts of the zeta function are obtained easily by taking advantage of the useful
fact \cite{Santos:1999yj} that, for a well behaved $G(s)$, the derivative of $G(s)/\Gamma(s)$ at $s=0$ is simply $G(0)$. 
I find
\begin{equation}
\zeta'_{HE,0}(0)={V\beta \over 8\pi^2}
\int_0^\infty dt \, t^{-3} eBt \coth eBt,
\label{zetae0prime}
\end{equation}
and its contribution to the free energy is
\begin{equation}
F_{HE,0}={V\over 16\pi^2}
\int_0^\infty dt \, t^{-3} eBt \coth eBt,
\label{Fe0prime}
\end{equation}
where we recognize immediately that $F_{HE,0}=-V{\cal L}_{HE,0}$, where ${\cal L}_{HE,0}$ is the unrenormalized
Heisenberg-Euler effective lagrangian for massless Majorana fermions. Similarly I find
\begin{equation}
\zeta'_{HE,\beta}(0)={V\beta \over 4\pi^2}
\int_0^\infty dt \, t^{-3} eBt \coth eBt
\left(\sum_{m=1}^\infty (-1)^m e^{-\beta^2 m^2/4t}\right),
\label{zetaebprime}
\end{equation}
whose contribution to the free energy is
\begin{equation}
F_{HE,\beta}={V\over 8\pi^2}
\int_0^\infty dt \, t^{-3} eBt \coth eBt
\left(\sum_{m=1}^\infty (-1)^m e^{-\beta^2 m^2/4t}\right),
\label{Febprime}
\end{equation}
and $F_{HE,\beta}=-V{\cal L}_{HE,\beta}$, where ${\cal L}_{HE,\beta}$ is the finite temperature correction
to the Heisenberg-Euler effective lagrangian \cite{elmfors2} for massless Majorana fermions. Notice that $F_{HE,0}$ and
$F_{HE,\beta}$ do not depend on the plates distance $a$ except for a multiplicative factor of $a$
contained in $V$.
The remaining two parts of the zeta function, $\zeta_{C,0}$ and $\zeta_{C,\beta}$, are the ones mostly contributing to 
the Casimir force, and their derivatives are given by
\begin{equation}
\zeta'_{C,0}(0)={V\beta \over 4\pi^2}
\int_0^\infty dt \, t^{-3} eBt \coth eBt
\left(\sum_{l=1}^\infty (-1)^l e^{-a^2 l^2/t}\right)
\label{zetac0prime}
\end{equation}
and
\begin{equation}
\zeta'_{C,\beta}(0)={V\beta \over 2\pi^2}
\int_0^\infty dt \, t^{-3} eBt \coth eBt
\left(\sum_{l=1}^\infty (-1)^l e^{-a^2 l^2/t}\right)
\left(\sum_{m=1}^\infty (-1)^m e^{-\beta^2 m^2/4t}\right).
\label{zetacbprime}
\end{equation}
Their contributions to the free energy are 
\begin{equation}
F_{C,0}={\zeta'_{C,0}(0)\over 2\beta} 
\label{FC0}
\end{equation}
and 
\begin{equation}
F_{C,\beta}={\zeta'_{C,\beta}(0)\over 2\beta} 
\label{FCb}
\end{equation}
respectively.
It is not possible to evaluate (\ref{zetacbprime}) in closed form for arbitrary values of $B$, $a$ and $\beta$, but it 
is possible to find simple expressions for (\ref{zetac0prime}) and (\ref{zetacbprime}) when the magnetic field is weak,
$eB\ll a^{-2}, \beta^{-2}$, and when $B$ is strong, $eB\gg a^{-2}, \beta^{-2}$.

In the case of weak magnetic field I can set 
\begin{equation}
eBt \coth eBt\approx 1+{1\over 3} (eBt)^2
\label{smallB}
\end{equation}
and, after substituting (\ref{smallB}) into (\ref{zetac0prime}), integrate by changing variable from 
$t$ to $1\over x$, to find
\begin{equation}
\zeta'_{C,0}(0)={V\beta \over 4\pi^2}\left[-{7\over 8}a^{-4}\zeta_R(4)\Gamma(2)+{(eB)^2\over 3}
{\partial f(s=0,a)\over\partial s}\right],
\label{zc0low}
\end{equation}
where $\zeta_R(4)=\pi^4/90$ is the Riemann zeta function of number theory, $\Gamma(2)=1$ is the Euler gamma function
and 
\begin{equation}
f(s,a)=a^{2s}(2^{2s+1}-1)\zeta_R(-2s){\Gamma(-s)\over \Gamma(s)}.
\label{f}
\end{equation}
I find
\begin{equation}
{\partial f(s=0,a)\over\partial s}=\gamma_E+2\zeta'_R(0)+\ln(4a),
\label{df}
\end{equation}
where $\gamma_E=0.5772$ is the EulerÐMascheroni constant and $\zeta'_R(0)=-0.9189$ is the derivative of the Riemann zeta function.
The interesting numerical fact 
\begin{equation}
\gamma_E+2\zeta'_R(0)+\ln 4\approx {1\over 8}
\label{numfact}
\end{equation}
allows me to write
\begin{equation}
{\partial f(s=0,a)\over\partial s}=\ln a + {1\over 8},
\label{df2}
\end{equation}
and, after inserting (\ref{df2}) into (\ref{zc0low}) and then (\ref{zc0low}) into (\ref{FC0}), I find
\begin{equation}
F_{C,0}=-\left({7\over 8}\right){\pi^2\over 720}{V\over a^4}+{(eB)^2\over 24\pi^2}V\left(\ln a +{1\over8}\right),
\label{FC0small}
\end{equation}
which, for $B=0$, is the vacuum Casimir energy for massless 
Majorana fermions with bag boundary conditions and
is in agreement with \cite{Johnson:1975zp}, and 
with \cite{Oikonomou:2009zr}, where the same result is obtained by a different method,
using a piston model to derive the vacuum Casimir energy for Majorana fermions. 
Notice that the second term in (\ref{FC0small}) is the
correction to the Casimir energy in vacuum due to the weak magnetic field.

In the case of weak magnetic field I insert (\ref{smallB}) into (\ref{zetacbprime}) and 
integrate as described previously, to obtain
\begin{equation}
\zeta'_{C,\beta}(0)={V\beta \over 2\pi^2}
\left[h(a,\beta)\Gamma(2)
+{(eB)^2\over 3}{\partial g(s=0,a,\beta)\over\partial s}\right],
\label{zetacbprime2}
\end{equation}
where the functions $g(s,a,\beta)$ and $h(a,\beta)$ are given by
\begin{equation}
g(s,a,\beta)=\left[(1+4^{s+1})E_2\left( -s; \,a^2, {\beta^2\over 4} \right)-2E_2\left( -s; \,a^2, \beta^2 \right)
-2E_2\left( -s; \,4a^2, {\beta^2\over 4}\right)\right]{\Gamma(-s)\over \Gamma(s)},
\label{g}
\end{equation}
\begin{equation}
h(a,\beta)={5\over 4}E_2\left( 2; \,a^2, {\beta^2\over 4} \right)-2E_2\left( 2; \,a^2, \beta^2 \right)
-2E_2\left( 2; \,4a^2, {\beta^2\over 4} \right),
\label{h}
\end{equation}
and I have expressed the double sum in terms of Epstein functions 
\cite{Santos:1999yj,Elizalde:1988rh,Kirsten:1994yp}
which for any positive integer $N$ are defined by
\begin{equation}
E_N\left( s; \,a_1, a_2,..., a_N\right)=\sum_{n_1=1}^\infty\sum_{n_2=1}^\infty....\sum_{n_N=1}^\infty
{1\over (a_1n_1^2 + a_2 n_2^2 +....+ a_Nn_N^2)^s}.
\label{Epstein}
\end{equation}
The Epstein functions can be analytically continued to meromorphic functions in the complex plane
\cite{Santos:1999yj,Elizalde:1988rh}, and for $N=2$ their analytic continuation is given by:
\begin{equation}
E_2\left( s; \,a_1, a_2\right)=-{a_1^{-s}\over 2}+{1\over 2}\sqrt{\pi\over a_2}
{\Gamma(s-{1\over 2})\over\Gamma(s)}E_1\left( s-{1\over 2}; \,a_1\right)+
{2\pi^s\over \Gamma(s) a_2^{{s\over 2}+{1\over 4}}}\sum_{n,m=1}^\infty 
\left({m\over\sqrt{a_1} n}\right)^{s-{1\over 2}}
K_{s-{1\over 2}}\left(2\pi m n\sqrt{a_1\over a_2}
\right),
\label{Epstein2}
\end{equation}
where $K_\nu(z)$ are modified Bessel functions. This analytic continuation of the Epstein
functions allows me to find
\begin{equation}
h(a,\beta)={7\pi^4\over 90}{1\over \beta^4}+{2\pi^2\over a^{5\over 2}\beta^{3\over 2}}
\sum_{n,m=1}^\infty 
\left(n\over m\right)^{3\over 2}\left[{5\over\sqrt{2}}K_{3\over 2}\left({\pi m n \beta\over a}\right)-
K_{3\over 2}\left({\pi m n \beta\over 2a}\right)-2K_{3\over 2}\left({2\pi m n \beta\over a}\right)
\right]
\label{h2}
\end{equation}
and
\begin{equation}
{\partial g(s=0,a,\beta)\over\partial s}=2\left({\beta\over a}\right)^{1\over 2}\sum_{n,m=1}^\infty 
\left(n\over m\right)^{1\over 2}\left[{5\over\sqrt{2}}K_{1\over 2}\left({\pi m n \beta\over a}\right)-
K_{1\over 2}\left({\pi m n \beta\over 2a}\right)-2K_{1\over 2}\left({2\pi m n \beta\over a}\right)
\right]-{1\over 2}\left[\ln \left({\beta\over 2}\right) +{1\over 8}\right],
\label{dg}
\end{equation}
where I used (\ref{numfact}). With the help of the following
\begin{equation}
K_{n+{1\over 2}}(z)=\sqrt{\pi\over 2z}\sum_{k=1}^n {(n+k)!\over k! (n-k)!(2z)^k}
\label{K}
\end{equation}
for the modified Bessel functions of half-integral order, I find
\begin{equation}
h(a,\beta)={7\pi^4\over 90}{1\over \beta^4}+{2\pi\over a\beta^3}\sum_{n,m=1}^\infty {1\over m^3}
\left[{5\over 2}\left(1+{\pi nm \beta\over a}\right)e^{-{\pi nm \beta\over a}}-
\left(2+{\pi nm \beta\over a}\right)e^{-{\pi nm \beta\over 2a}}
-\left({1\over 2}+{\pi nm \beta\over a}\right)e^{-{2\pi nm \beta\over a}}
\right],
\label{h3}
\end{equation}
and
\begin{equation}
{\partial g(s=0,a,\beta)\over\partial s}=2\sum_{n,m=1}^\infty{1\over m}
\left[{5\over 2}e^{-{\pi nm \beta\over a}}-
e^{-{\pi nm \beta\over 2a}}-e^{-{2\pi nm \beta\over a}}
\right]
-{1\over 2}\left[\ln \left({\beta\over 2}\right) +{1\over 8}\right].
\label{dg2}
\end{equation}
From (\ref{h3}) and (\ref{dg2}) I can easily obtain the low temperature expression of $h$ and 
${\partial g\over\partial s}$. For $aT\ll 1$, I find
\begin{equation}
h(a,\beta)={7\pi^4\over 90}{1\over \beta^4}-{2\pi^2\over a^2\beta^2}
\left(1+{2a\over \pi\beta}\right)e^{-{\pi \beta\over 2a}},
\label{hlowT}
\end{equation}
and
\begin{equation}
{\partial g(s=0,a,\beta)\over\partial s}=-2e^{-{\pi  \beta\over 2a}}
-{1\over 2}\left[\ln \left({\beta\over 2}\right) +{1\over 8}\right].
\label{dglowT}
\end{equation}
I insert (\ref{hlowT}) and (\ref{dglowT}) into (\ref{zetacbprime2}), then  (\ref{zetacbprime2})
into (\ref{FCb}) and find immediately the finite 
temperature correction to the Casimir energy in the case of weak magnetic field 
and low temperature. For $eB\ll T^2\ll a^{-2}$, I obtain
\begin{equation}
F_{C,\beta}=
{7\pi^2\over 360}{V\over \beta^4}-{V\over 2a^2\beta^2}
\left(1+{2a\over \pi\beta}\right)e^{-{\pi \beta\over 2a}}
-{(eB)^2\over 24\pi^2}V\left[4e^{-{\pi  \beta\over 2a}}
+\ln \left({\beta\over 2}\right) +{1\over 8}\right].
\label{FCb2}
\end{equation}
In order to calculate the Helmholtz free energy, I need to add the contribution of the Heisenberg-Euler 
effective lagrangian (\ref{Fe0prime}) and of its finite temperature correction (\ref{Febprime}) 
to the Casimir energy and its thermal correction. 
In the case of weak magnetic field, $eB\ll T^2$, these contributions are \cite{Dittrich:1979ux,Elmfors:1993wj,Erdas:2010yq}
\begin{equation}
F_{HE,0}+F_{HE,\beta}=-{7\pi^2\over 360}
{V\over \beta^4}+{V(eB)^2\over 48\pi^2}\left[
\ln \left({\beta^2 eB\over 4}\right)+{1\over 4}\right],
\label{FHE}
\end{equation}
and, once I add (\ref{FHE}) to the vacuum Casimir energy for weak magnetic field (\ref{FC0small}) and to its
finite temperature correction (\ref{FCb2}), I obtain the Helmholtz free energy $F$ in the limit $eB\ll T^2\ll a^{-2}$
\begin{equation}
F=-\left({7\over 8}\right){\pi^2\over 720}{V\over a^4}-{V\over 2a^2\beta^2}
\left(1+{2a\over \pi\beta}\right)e^{-{\pi \beta\over 2a}}
-{(eB)^2\over 24\pi^2}V\left[4e^{-{\pi  \beta\over 2a}}
-{1\over 2}\ln \left(eBa^2\right)\right].
\label{FC}
\end{equation}
Notice the cancellation of the Stefan-Boltzmann term and of the temperature dependence from the logarithmic term.
Due to these cancellations, I find that the leading order correction to the Casimir vacuum energy is the temperature independent
and magnetic field dependent logarithmic term which, even in the case of weak magnetic field, is larger than the exponentially 
suppressed part, for $Ta\ll 1$. The pressure $P$ on the plates is given by
\begin{equation}
P=-{1\over L^2}{\partial F\over\partial a},
\label{P}
\end{equation}
and therefore, in the limit $eB\ll T^2\ll a^{-2}$, I find
\begin{equation}
P=-\left({7\over 8}\right){\pi^2\over 240 a^4}+{\pi\over 4a^3\beta}
e^{-{\pi \beta\over 2a}}
+{(eB)^2\over 12\pi^2}\left[{\pi\beta\over a}e^{-{\pi  \beta\over 2a}}
-{1\over 4}\ln \left(eBa^2\right)\right].
\label{P2}
\end{equation}

Another interesting limit, for the weak field case, is the high temperature case of $eB\ll a^{-2}\ll T^2$. Exploiting the fact
that (\ref{zetacbprime}) is invariant if I exchange $a$ and $\beta/2$, I write
$h$ and $g$ in a different form, which is better suited for the high temperature expansion, $aT\gg 1$:
\begin{eqnarray}
h(a,\beta)=&&\left({7\over8}\right){\pi^4\over 180}{1\over a^4}+{\pi\over 2a^3\beta}\sum_{n,m=1}^\infty {1\over m^3}
\nonumber \\
&&\times\left[{5\over 2}\left(1+{4\pi nm a\over \beta}\right)e^{-{4\pi nm a\over \beta}}-
\left(2+{4\pi nm a\over \beta}\right)e^{-{2\pi nm a\over \beta}}
-\left({1\over 2}+{4\pi nm a\over \beta}\right)e^{-{8\pi nm a\over \beta}}
\right]
\label{h4}
\end{eqnarray}
and
\begin{equation}
{\partial g(s=0,a,\beta)\over\partial s}=2\sum_{n,m=1}^\infty{1\over m}
\left({5\over 2}e^{-{4\pi nm a\over \beta}}-
e^{-{2\pi nm a\over \beta}}-e^{-{8\pi nm a\over \beta}}
\right)
-{1\over 2}\left(\ln a +{1\over 8}\right),
\label{dg3}
\end{equation}
from which I find immediately the high temperature expansions of $h$ and $\partial g
\over \partial s$
\begin{equation}
h(a,\beta)={7\pi^4\over 1,440}{1\over a^4}-{\pi\over a^3\beta}
\left(1+{2\pi a\over \beta}\right)e^{-{2\pi a\over \beta}}
\label{h5}
\end{equation}
and
\begin{equation}
{\partial g(s=0,a,\beta)\over\partial s}=-2
e^{-{2\pi a\over \beta}}
-{1\over 2}\left(\ln a +{1\over 8}\right).
\label{dg5}
\end{equation}
I use the last two equations to obtain the high temperature limit of $F_{C,\beta}$
\begin{equation}
F_{C,\beta}=\left({7\over 8}\right){\pi^2\over 720}{V\over a^4}
-{1\over 2}{V\over a^2\beta^2}\left(1+{\beta\over 2 \pi a}\right)e^{-{2\pi a\over \beta}}
-{(eB)^2\over 24\pi^2}V\left(4e^{-{2\pi a\over \beta}}
+\ln a +{1\over 8}\right).
\label{FCbhigh}
\end{equation}
Adding (\ref{FCbhigh}) to (\ref{FC0small}) and (\ref{FHE}), I obtain the free energy in the weak field and 
high temperature limit $eB\ll a^{-2}\ll T^2$
\begin{equation}
F=-\left({7\over 8}\right){\pi^2\over 45}{V\over \beta^4}
-{1\over 2}{V\over a^2\beta^2}\left(1+{\beta\over 2 \pi a}\right)e^{-{2\pi a\over \beta}}
-{(eB)^2\over 24\pi^2}V\left[4e^{-{2\pi a\over \beta}}
-{1\over 2}\ln \left({\beta^2eB\over 4}\right) -{1\over 8}\right].
\label{Fhigh}
\end{equation}
Notice that the leading term here is the Stefan-Boltzmann term. 
The pressure in the weak field and 
high temperature limit is obtained immediately from (\ref{Fhigh})
\begin{equation}
P=\left({7\over 8}\right){\pi^2\over 45}{1\over \beta^4}-
{\pi\over a \beta^3}e^{-{2\pi a\over \beta}}
-{(eB)^2\over 24\pi^2}\left[{8a\over\beta}e^{-{2\pi a\over \beta}}
+{1\over 2}\ln \left({\beta^2eB\over 4}\right) +{1\over 8}\right],
\label{P3}
\end{equation}
where the dominant piece is clearly the outward pressure originating from the Stefan-Boltzmann term of the free energy.

In the case of strong magnetic field  
\begin{equation}
eBt \coth eBt\approx eBt
\label{highB}
\end{equation}
and, after substituting (\ref{highB}) into (\ref{zetac0prime}) and integrating, 
I find
\begin{equation}
\zeta'_{C,0}(0)=-{V\beta \over 8\pi^2}{eB\over a^2}\Gamma(1)\zeta_R(2).
\label{zc0high}
\end{equation}
From (\ref{zc0high}) I obtain immediately the vacuum piece of the Casimir energy for strong magnetic field
\begin{equation}
F_{C,0}=-{V\over 96}{eB\over a^2}.
\label{Fc0high}
\end{equation}

In order to find the thermal correction to $F_{C,0}$, I substitute (\ref{highB}) into (\ref{zetacbprime}), 
integrate and obtain
\begin{equation}
\zeta'_{C,\beta}(0)={V\beta \over \pi^2}eB
\left[E_2\left(1;a^2,{\beta^2\over4}\right)-E_2\left(1;a^2,{\beta^2}\right)
-E_2\left(1;4a^2,{\beta^2\over4}\right)\right],
\label{zetacbprime3}
\end{equation}
which, using the analytic continuation (\ref{Epstein2}) of the Epstein functions and the expression of the 
modified Bessel functions (\ref{K}), can be written as
\begin{equation}
\zeta'_{C,\beta}(0)=V eB
\left[{1\over 12\beta}+{2\over\pi a}\sum_{n,m=1}^\infty{1\over m}
\left(e^{-{\pi nm \beta\over a}}-
e^{-{\pi nm \beta\over 2a}}-{1\over 2}e^{-{2\pi nm \beta\over a}}\right)\right].
\label{zetacbprime4}
\end{equation}
From (\ref{zetacbprime4}) I obtain $F_{C,\beta}$, the thermal correction to the Casimir energy
in the case of strong magnetic field
\begin{equation}
F_{C,\beta}=
{VeB\over 24\beta^2}+{VeB\over\pi a\beta}\sum_{n,m=1}^\infty{1\over m}
\left(e^{-{\pi nm \beta\over a}}-
e^{-{\pi nm \beta\over 2a}}-{1\over 2}e^{-{2\pi nm \beta\over a}}\right).
\label{zetacbprime5}
\end{equation}
I now use this equation to obtain the low temperature limit of $F_{C,\beta}$ and,
for $eB \gg a^{-2}\gg T^2$, find
\begin{equation}
F_{C,\beta}=
{VeB\over 24\beta^2}-{VeB\over\pi a\beta}
e^{-{\pi \beta\over 2a}}.
\label{zetacbprime6}
\end{equation}
In the case of strong magnetic field, $eB\gg T^2$, the contribution of the Heisenberg-Euler 
effective lagrangian and of its finite temperature correction
to the free energy is \cite{Dittrich:1979ux,Elmfors:1993wj,Erdas:2010yq}
\begin{equation}
F_{HE,0}+F_{HE,\beta}=-{V\over 24}{eB\over \beta^2}-{V(eB)^2\over 48\pi^2}
\ln \left({\beta^2 eB}\right). 
\label{FHE2}
\end{equation}
I add (\ref{FHE2}), (\ref{zetacbprime6}) and (\ref{Fc0high}) and obtain the free energy $F$ and the 
pressure $P$, in the limit
$eB \gg a^{-2}\gg T^2$
\begin{equation}
F=-{V(eB)^2\over 48\pi^2}\ln \left({\beta^2 eB}\right)-{V\over 96}{eB\over a^2}
-{VeB\over\pi a\beta}
e^{-{\pi \beta\over 2a}},
\label{Fhigh2}
\end{equation}
\begin{equation}
P={(eB)^2\over 48\pi^2}\ln \left({\beta^2 eB}\right)-{1\over 96}{eB\over a^2}
+{eB\over 2a^2}
e^{-{\pi \beta\over 2a}}.
\label{P4}
\end{equation}
In both of these equations, the dominant term is quadratic in the magnetic field and comes
from the Heisenberg-Euler effective lagrangian.

Finally I want to find $F$ and $P$ in the case of strong magnetic field and high temperature,
$eB \gg T^2\gg a^{-2}$. I again exploit the symmetry of (\ref{zetacbprime}) for
exchange of $a$ and $\beta/2$ to write (\ref{zetacbprime5}) in an equivalent form, which is better suited for 
expansion in the case of $T^2\gg a^{-2}$
\begin{equation}
F_{C,\beta}=
{VeB\over 96a^2}+{VeB\over\pi a\beta}\sum_{n,m=1}^\infty{1\over m}
\left(e^{-{4\pi nm a\over \beta}}-
e^{-{2\pi nm a\over \beta}}-{1\over 2}e^{-{8\pi nm a\over \beta}}\right),
\label{FCb6}
\end{equation}
and its high temperature limit is
\begin{equation}
F_{C,\beta}=
{VeB\over 96a^2}-{VeB\over\pi a\beta}
e^{-{2\pi a\over \beta}}.
\label{FCb7}
\end{equation}
By adding (\ref{FCb7}) to (\ref{zetacbprime6}) and (\ref{Fc0high}) I obtain the free energy and the pressure
in the case of strong magnetic field and high temperature
\begin{equation}
F=-{V(eB)^2\over 48\pi^2}\ln \left({\beta^2 eB}\right)-{V\over 24}{eB\over \beta^2}
-{VeB\over\pi a\beta}
e^{-{2\pi a\over \beta}},
\label{Fhigh3}
\end{equation}
\begin{equation}
P={(eB)^2\over 48\pi^2}\ln \left({\beta^2 eB}\right)+{1\over 24}{eB\over \beta^2}
-{2eB\over \beta^2}
e^{-{2\pi a\over \beta}}.
\label{P5}
\end{equation}
\section{Discussion and conclusions}
\label{4}
In this paper I used the zeta function regularization technique to study the finite
temperature Casimir effect
of a massless Majorana fermion field confined between parallel plates and in the 
presence of a magnetic field perpendicular to the plates. I have obtained an expression 
 for the zeta function (\ref{zeta3}) which is exact to all orders in the magnetic 
field strength $B$, and from it I have derived expressions for the Helmholtz free
energy and for the pressure on the plates in the case of weak magnetic field
($eB\ll a^{-2}, T^2$) and in the case of strong magnetic field
($eB\gg a^{-2}, T^2$).

In the case of a weak magnetic field, I found the 
temperature correction to the Casimir energy to be
\begin{equation}
F_{C,\beta}={V \over 4\pi^2}
\left[h(a,\beta)
+{(eB)^2\over 3}{\partial g(s=0,a,\beta)\over\partial s}\right],
\label{D01}
\end{equation}
and obtained two equivalent expressions for $h$, (\ref{h3}) and
(\ref{h4}), which are exact to all orders in $a$ and $\beta$, and two
equivalent expressions for $\partial g\over \partial s$, (\ref{dg2}) 
and (\ref{dg3}), which are also exact to all orders in $a$ and $\beta$. These expressions
involve two infinite double sums, and I have been able to evaluate them numerically
with very high precision. I find that, for $0\le aT \le {1\over 2}$, the simple low temperature expression of $h$ 
that I obtained in Eq. (\ref{hlowT}) is within less than one percent of the exact value of $h$, as given 
by the infinite double sum of Eq. (\ref{h3}). Similarly I find that, for ${1\over 2}\le aT \le \infty$, the high
temperature expression of $h$ that I write in Eq. (\ref{h5}) is within less than one percent of the exact
value of $h$. A similar numerical evaluation of the exact expression of $\partial g\over \partial s$ leads
me to discover that the low temperature expression of $\partial g\over \partial s$, (\ref{dglowT}), is 
within less than four percent of its exact value in the range $0\le aT \le {1\over 2}$, while 
the high temperature  expression of $\partial g\over \partial s$, (\ref{dg5}), is 
within less than four percent of its exact value in the range ${1\over 2}\le aT \le \infty$. I summarize these findings
by writing the free energy $F$ for weak magnetic field as
\begin{equation}
F=\cases{
-\left({7\over 8}\right){\pi^2\over 720}{V\over a^4}-{V\over 2a^2\beta^2}
\left(1+{2a\over \pi\beta}\right)e^{-{\pi \beta\over 2a}}
-{(eB)^2\over 24\pi^2}V\left[4e^{-{\pi  \beta\over 2a}}
-{1\over 2}\ln \left(eBa^2\right)\right]
& \text{for $0\le aT \le {1\over 2}$ ;}
\cr
-\left({7\over 8}\right){\pi^2\over 45}{V\over \beta^4}
-{V\over 2a^2\beta^2}\left(1+{\beta\over 2 \pi a}\right)e^{-{2\pi a\over \beta}}
-{(eB)^2\over 24\pi^2}V\left[4e^{-{2\pi a\over \beta}}
-{1\over 2}\ln \left({\beta^2eB\over 4}\right) \right]
& 
\text{for ${1\over 2}\le aT \le \infty$ .}
\cr}
\label{Ffinal}
\end{equation}
When evaluated at $aT={1\over 2}$, both expressions of $F$ yield the same value. Equation (\ref{Ffinal}) is a simple analytic expression 
of $F$ for weak magnetic field, valid for all values of 
the temperature $T$ and the plate distance $a$, and with a discrepancy of no more than a few percent from 
the exact value of $F$. A similarly accurate expression of the pressure $P$, valid for weak magnetic field and all values of $a$ and $T$,
is obtained immediately from (\ref{Ffinal}), since $P=-{1\over L^2}{\partial F\over\partial a}$. 
Notice that, if we set $B=0$ in (\ref{Ffinal}), we obtain the free energy for the finite temperature Casimir effect of Majorana fermion fields and,
opposite to what is claimed in \cite{Cheng:2010kc}, the free energy is always negative and does not become positive for $aT>0.37$.
Even if we do not include the contribution of the thermal part of the 
effective Lagrangian and consider only $F_{C,0}+F_{C,\beta}$
at $B=0$, we find
\begin{equation}
F_{C,0}+F_{C,\beta}=\cases{
-\left({7\over 8}\right){\pi^2\over 720}{V\over a^4}
+\left({7\over 8}\right){\pi^2\over 45}{V\over \beta^4}
-{V\over 2a^2\beta^2}
\left(1+{2a\over \pi\beta}\right)e^{-{\pi \beta\over 2a}}
& \text{for $0\le aT \le {1\over 2}$ ;}
\cr
-{V\over 2a^2\beta^2}\left(1+{\beta\over 2 \pi a}\right)e^{-{2\pi a\over \beta}}
& 
\text{for ${1\over 2}\le aT \le \infty$ ,}
\cr}
\label{Ffinal01}
\end{equation}
which is negative for $0\le aT < \infty$.

In the case of a strong magnetic field, two equivalent expressions of the exact  
temperature correction to the Casimir energy are presented in Eqs. (\ref{zetacbprime5}) and (\ref{FCb6}), both involving
two infinite double sums. After a highly precise numerical evaluation of these two expressions, I find that the low temperature
limit of $F_{C,\beta}$ presented in (\ref{zetacbprime6}), is accurate within less than two percent for $0\le aT \le {1\over 2}$, while the high temperature 
limit of $F_{C,\beta}$ shown in (\ref{FCb7}) is also accurate  within less than two percent for ${1\over 2}\le aT \le \infty$.
These findings
allow me to write the free energy for strong magnetic field as
\begin{equation}
F=\cases{
-{V(eB)^2\over 48\pi^2}\ln \left({\beta^2 eB}\right)-{V\over 96}{eB\over a^2}
-{VeB\over\pi a\beta}
e^{-{\pi \beta\over 2a}}& \text{for $0\le aT \le {1\over 2}$ ;}
\cr
-{V(eB)^2\over 48\pi^2}\ln \left({\beta^2 eB}\right)-{V\over 24}{eB\over \beta^2}
-{VeB\over\pi a\beta}
e^{-{2\pi a\over \beta}},
& 
\text{for ${1\over 2}\le aT \le \infty$ ,}
\cr}
\label{Ffinal2}
\end{equation}
a simple analytic expression of $F$, valid for all values of 
$T$ and $a$, and with a discrepancy of no more than one or two percent from 
the exact value of $F$. The pressure in the case of strong magnetic field is obtained immediately from
(\ref{Ffinal2}) for all values of $a$ and $T$.

\end{document}